\documentclass[conference]{IEEEtran}

\usepackage{cite}
\usepackage{url}

\usepackage{graphicx}
\usepackage{algorithm}
\usepackage[noend]{algorithmic}
\usepackage{subfig}
\usepackage{amssymb, amsmath,graphicx,charter, latexsym}

\newtheorem{theorem}{Theorem}

\def\baselinestretch{0.85}
\begin{document}
\title{Distributed Resource Allocation for Proportional Fairness in Multi-Band Wireless Systems}
\author{
\IEEEauthorblockN{I-Hong Hou} \IEEEauthorblockA{CSL and Department
of CS\\University of Illinois\\Urbana, IL 61801,
USA\\ihou2@illinois.edu} \and \IEEEauthorblockN{Piyush
Gupta}\IEEEauthorblockA{Bell Labs, Alcatel-Lucent\\600 Mountain
Ave., 2C-374\\Murray Hill, NJ 07974\\pgupta@research.bell-labs.com}
 }
\maketitle

\begin{abstract}
A challenging problem in multi-band multi-cell self-organized
wireless systems, such as multi-channel Wi-Fi networks, femto/pico
cells in 3G/4G cellular networks, and more recent wireless networks
over TV white spaces, is of distributed resource allocation. This
involves four components: channel selection, client association,
channel access, and client scheduling. In this paper, we present a
unified framework for jointly addressing the four components with
the global system objective of maximizing the clients throughput in
a proportionally fair manner. Our formulation allows a natural
dissociation of the problem into two sub-parts. We show that the
first part, involving channel access and client scheduling, is
convex and derive a distributed adaptation procedure for achieving
Pareto-optimal solution. For the second part, involving channel
selection and client association, we develop a Gibbs-sampler based
approach for local adaptation to achieve the global objective, as
well as derive fast greedy algorithms from it that achieve good
solutions.
\end{abstract}

\section{Introduction}  \label{section:introduction}
Many of the existing and evolving wireless systems operate over
spectrum that spans multiple bands. These bands may be contiguous,
as in OFDM-based systems, such as current IEEE 802.11-based WLANs
(a.k.a.~Wi-Fi networks) and evolving fourth-generation LTE cellular
wireless systems; or they may be spread far apart, as in
multi-channel 802.11 systems and in recently proposed wireless
broadband networks over TV white spaces (discussed in the sequel). A
common issue in these multi-band systems is how to perform resource
allocation among different clients, possibly being served by
different access points (APs). This needs to be done so as to
efficiently utilize wireless resources---spectrum, transmission
opportunities and power---while being fair to different clients.
Furthermore, unlike traditional enterprise Wi-Fi networks and
cellular wireless networks, where the placement of APs and their
operating bands are arrived at after careful capacity/coverage
planning, more and more of the evolving wireless systems are going
to be self-organized networks. There is extensive literature on
completely self-organized wireless networks, also referred to as
{\em ad hoc} networks \cite{toh02}, which are often based on 802.11.
Even the emerging 4G cellular wireless networks, such as those based
on LTE, are going to have a significant deployment of self-organized
subsystems: namely, pico cells and femto cells \cite{chaandgat08}.
These self-organized (sub)-systems will require that resource
allocation is performed dynamically in a distributed manner and with
minimal coordination between different APs and/or clients.

Another emerging scenario is of broadband wireless networks
operating over TV white spaces \cite{bahchamos09,debsrimha09}.
Recent conversion to all-digital TV broadcast has made available
valuable lower-frequency spectrum. A part of this spectrum, referred
to as TV white spaces, has been mandated by FCC for unlicensed
broadband access.  However, designing a networking stack over the
large and fragmented TV white spaces, which may have widely
different propagation characteristics, pose new technical challenges
unseen in traditional wireless networks and call for entirely new
wireless design principles.

In this paper, we consider the problem of joint resource allocation
across different APs and their clients so as to achieve a global
objective of maximizing the system throughput while being fair to
users. Unlike the max-min fairness used in traditional Wi-Fi
networks, which has been shown in several extensive studies to be
inefficient \cite{heurouber03,tangut04}, we will focus on
proportional fairness. The latter has become essentially standard
across current 3G cellular systems, as well as in emerging 4G
systems based on LTE and WiMAX. Thus, the system objective will be
to allocate wireless resources, spectrum and transmission
opportunities, so as to maximize the (weighted) sum of the log of
throughputs of different clients, which is known to achieve
(weighted) proportional fairness.

The combined resource allocation in a multi-band multi-cell wireless
system involves four components: Channel selection, client
association,  channel access, and client scheduling. The first
component, Channel Selection, decides on how different APs share
different bands of the spectrum available to the wireless system. As
mentioned earlier, the emphasis will be on self-organized networks,
which will necessitate a completely distributed approach that
dynamically adapts to varying traffic requirements in different
cells. To achieve these objectives, we will consider a
differentiated random-access based solution. The second component,
Client Association, allows a client to decide on an AP to associate
within its neighborhood that is likely to provide the ``best"
performance. Unlike the traditional approach of associating with the
AP having the highest signal strength, we will discuss an approach
that allows a client to associate with an AP that maximizes its
proportionally-fair throughput while minimizing the impact on other
clients. Once an AP has chosen a channel/band to operate in and a
bunch of clients have associated with it, the third component,
Channel Access, decides when it should access the channel so as to
serve its clients while being fair to other access points in its
neighborhood operating in the same channel. The final component,
Client Scheduling, decides which of its clients an AP should serve
whenever it successfully accesses the channel.

Our approach addresses the four components in a unified framework,
where the solutions to different components are arrived at through
separation of time scales of adaptation. More specifically, our
formulation allows for the optimization problem of maximizing the
clients throughput with weighted proportional fairness  to naturally
dissociate into two sub-parts, which are adapted at different time
scales. Assuming that the channel selection and client association
have been performed, we show that the sub-problem of channel access
and client scheduling becomes convex, which is also amenable to a
distributed adaptation for achieving Pareto-optimal weighed
proportional fairness. At a slower time scale, we adapt the channel
selection and client association to varying demands and
interference. This part is a non-convex problem in general, and
thus, difficult to solve for a globally optimal solution. We develop
a Gibbs-sampler based approach to perform local adaptation while
improving the global system objective. The adaptation is randomized,
and if done slowly enough, can achieve a globally optimal solution.
In practice, however, that may not be always feasible; hence, we
derive greedy heuristics from it for channel selection and client
association, which though not globally optimal, provide fast and
good distributed solutions with limited exchange of information, as
the simulation results indicate.

The paper is organized as follows. Section~\ref{section:related
work} provides an overview of some related work.
Section~\ref{section:system model} describes the multi-cell
multi-band wireless system model and the joint resource allocation
problem that we study. Section~\ref{section:solution overview} gives
an overview of our approach and discusses the separation of
problems. The details of our approach and its desirable properties,
including convergence to Pareto-optimal proportionally-fair
allocation, are given in Section~\ref{section:scheduling contention}
and Section~\ref{section:AP channel}. The results of simulations are
provided in Section~\ref{section:simulation}. Concluding remarks are
discussed in Section~\ref{section:conclusion}.

\section{Related Work}  \label{section:related work}
There is extensive literature addressing one or a subset of the
aforementioned four components of wireless resource allocation in
the context of different wireless systems---Wi-Fi networks, 3G/4G
cellular networks, and more recent, wireless broadband systems over
TV white spaces. Due to space limitation, we only discuss a small
sample of the results in each of these areas.

Distributed resource allocation has been widely studied for IEEE
802.11-based systems.  A number of  CSMA-based random-access
approaches have been developed to provide differentiated services to
clients \cite{denhaa98, rozkum01, vercambar01,bacblamuh03,
gupsansto05}. Proportional fairness in multi-contention neighborhood
has been studied in \cite{karsartas04, gupsto06}. Specifically,
\cite{gupsto06} has established that Pareto-optimal weighted
proportional fairness can be achieved in a distributed manner with
minimal exchange of information among contending clients. Our
approach for channel access is motivated by \cite{gupsto06}; we,
however, consider a more general framework that jointly addresses
all the four components. Multi-channel MACs for Wi-Fi networks have
been proposed in \cite{bahchadun04,sovai04}. Approximation
algorithms for client association control to achieve proportional
fairness have been developed in \cite{bejhanli04,lipalyan08}, but
they assume that APs do not interfere with each other through a
pre-assignment of orthogonal channels. The work closest to this
paper is \cite{kaubaccha07} where the authors have developed
Gibbs-sampler based distributed algorithms for channel selection and
client association. Their approach, however, considers different
objectives for the two components, neither of which ensures
proportional fairness.

For cellular wireless data systems, a number of centralized
approaches for single-cell scheduling to achieve proportional
fairness have been developed \cite{benblagro00, andkumram01,
andqiasto05}. More recently, several inter-cell interference
coordination (ICIC) techniques have been proposed for interference
mitigation in LTE-based 4G cellular networks. Specifically,
\cite{stovis08,stovis09} have developed distributed algorithms for
dynamic fractional frequency reuse among interfering macro cells
based on limited exchange of interference information over dedicated
control links. Given the non-convexity of the problem, these
algorithms aim to achieve a local optimum of the weighted sum of the
log of user throughputs, which as mentioned earlier provides
proportional fairness, through local estimates of the appropriate
gradients.  However, in self-organized subsytems of such networks
(such as those made of pico and femto cells), explicit exchange of
information may not be feasible \cite{andcapfek10}, and thus,  these
approaches may not be directly applicable.

Research on resource allocation in wireless networks operating over
fragmented TV white spaces is in a nascent stage. \cite{bahchamos09}
considers a single AP serving multiple clients at the same rate. For
this scenario, it addresses three issues: how the AP chooses a
suitable band, how a new client detects the AP's operating band, and
how disruptions due to temporal variations, such as caused by
wireless microphones, are handled. Some of the limitations of this
approach are overcome in \cite{debgupkan10} by considering a
multi-rate multi-radio architecture. It develops a joint strategy
for white space selection and client assignment to one of the
radios, as well as designs an extension to CSMA to achieve
proportional fairness---all for a single-AP scenario. This paper
generalizes the setup by considering a system of multiple APs and
jointly addressing the four components of resource allocation to
achieve the global system objective of maximizing clients throughput
in a proportionally fair manner.

\section{System Model}  \label{section:system model}

We consider a system with several access points (APs) and clients
that can operate in a number of channels. We denote the set of APs
by $\mathbb{N}$, the set of clients by $\mathbb{I}$, and the set of
channels by $\mathbb{C}$. Each client $i$ is associated with an AP,
which we denote by $n(i)$, and is served by that AP. Each AP $n$ is
equipped with $u_n$ radios that can operate in different channels
simultaneously and each client $i$ is equipped with only one radio.
When an AP $n$ has more than one radios, i.e, $u_n>1$, we can
simplify the model by assuming that there are $u_n$ APs, each with
one radio, that are placed at the same place as AP $n$. Each of
these $u_n$ APs corresponds to one radio of AP $n$. This procedure
allows us to only consider the model in which each AP has one radio
and simplifies notations. Throughout the rest of the paper, we
assume that each AP only has one radio and operates in one channel
unless otherwise specified. The channel that an AP $n$ is operating
in is denoted by $c(n)$. APs can switch the channels that they
operate in, although such switches can only be done infrequently due
to the large overheads. When an AP switches channels, all its
clients also switch channels accordingly.

We focus on a server-centric scheme where each AP schedules all
transmissions between itself and all clients that are associated
with it. This scheme is applicable to a wide varieties of wireless
systems that include LTE, WiMax, and IEEE 802.11 PCF. Later, we will
also discuss a distributed scheme where clients contend for service
from APs, such as in 802.11 DCF. We assume that time is slotted,
with the duration of a time slot equals the time needed for a
transmission. If an AP $n$ makes a successful transmission toward
client $i$ in channel $c$, client $i$ receives data at rate of
$B_{i,n,c}$ in this slot. Since characteristics of different
channels may be different, $B_{i,n,c}$ depends on $c$.

On the other hand, we assume that the APs are not synchronized and
may interfere with each other. We consider the interference
relations using the protocol model \cite{GK2000}. When an AP $n$
operates in channel $c$, it may be interfered by a subset
$\mathcal{M}^{n,c}$ of APs, where $n\in \mathcal{M}^{n,c}$ for
notational simplicity. When the AP $n$ schedules a transmission
between itself and one of its clients, the transmission is
successful if $n$ is the only AP among $\mathcal{M}^{n,c}$ that
transmits in channel $c$ during the time of transmission; otherwise,
the transmission suffers from a collision and fails. We assume that
the interference relations are symmetric, i.e., $m\in
\mathcal{M}^{n,c}$ if and only if $n\in\mathcal{M}^{m,c}$. Note
that, since the propagation characteristics of different channels
may be different, especially in the case of TV white space access,
the subset $\mathcal{M}^{n,c}$ depends on the channel $c$. This
dependency further distinguishes our work from most existing works
on multi-channel access where interference relations are assumed to
be identical for all channels. We also define
$\mathcal{M}_{i}:=\{m\in\mathbb{N}|\exists c\in\mathbb{C},
n\in\mathcal{N}_{i} \mbox{ s.t. } m\in\mathcal{M}^{n,c}\}$ for each
client $i$.

Since APs are not coordinated, we assume that they access the
channel by random access. Each AP $n$ chooses a random access
probability, $p_n$. In each slot, AP $n$ accesses the channel $c(n)$
with probability $p_n$. The transmission is successful if $n$ is the
only AP in $\mathcal{M}^{n,c(n)}$ that transmits over channel
$c(n)$. Thus, the probability that AP $n$ successfully accesses the
channel can be expressed as
\begin{equation}
\label{equation:system model:access prob} \mbox{$
p_n\prod_{\hspace{-5pt}\begin{array}{l}\vspace{-5pt}^{m\in\mathcal{M}^{n,c(n)},}\\
^{m\neq n,c(m)=c(n)}\end{array}}\hspace{-10pt}(1-p_m)=\frac{p_n}{1-p_n}\prod_{\hspace{-5pt}\begin{array}{l}\vspace{-5pt}^{m\in\mathcal{M}^{n,c(n)},}\\
^{c(m)=c(n)}\end{array}}(1-p_m). $}
\end{equation}

The AP is in charge of scheduling transmissions for its clients.
When AP $n$ accesses the channel, it schedules the transmission for
client $i$, where $n(i)=n$, with probability $\phi_{i,n}$,
$\phi_{i,n}\geq0$ and $\sum_{i:n(i)=n}\phi_{i,n}=1$. Since the data
rate of client $i$ when it is served is $B_{i,n(i),c(n(i))}$ and the
probability that the AP $n(i)$ makes a successful transmission is as
in Eq. (\ref{equation:system model:access prob}), its throughput per
time slot is, assuming $n(i)=n$ and $c(n)=c$,
\begin{equation}
\label{equation:system model:rate}
\mbox{$r_i:=B_{i,n,c}\phi_{i,n}\frac{p_{n}}{1-p_{n}}\prod_{m\in\mathcal{M}^{n,c},c(m)=c}(1-p_m).$}
\end{equation}

We now discuss an analog model for a distributed scheme where
clients contend for the service from APs, which can be applied to
completely distributed scenarios, such as those based on 802.11 DCF.
In this scheme, each client $i$ contend for the channel by accessing
it with probability $p_i$ in each time slot. Two clients, $i$ and
$j$, interfere with each other if their associated APs interfere
with each other, that is, $c(n(i))=c(n(j))$ and $n(i)\in
\mathcal{M}^{n(j),c(n(j))}$. Client $i$ successfully accesses the
channel if none of the other clients that can interfere with it
access the channel simultaneously. Thus, the long-term throughput
per time slot for client $i$ is , assuming $n(i)=n$ and $c(n)=c$,
$r_i:=B_{i,n,c}\frac{p_i}{1-p_i}\prod_{j:c(n(j))=c, n(j)\in
\mathcal{M}^{n,c}}(1-p_j).$

Finally, we assume that each client is associated with a positive
weight $w_i>0$. We also denote the total weights of clients
associated with AP $n$ by $w^n$, i.e., $w^n:=\sum_{i:n(i)=n}w_i$.
The goal is to achieve weighted proportional fairness among clients,
that is, to maximize $\sum_{i\in\mathbb{I}}w_i\log r_i$. The
solution to the distributed scheme is very similar to that to the
server-centric scheme. Thus, we will focus on the server-centric
scheme and only report key results of the distributed scheme.

\section{Solution Overview and Time-Scale Separation}   \label{section:solution overview}

We now give an overview of our approach to achieve weighted
proportional fairness, which consists of separating the problem into
four components and solving them. By Eq. (\ref{equation:system
model:rate}), we can formulate the problem of achieving weighted
proportional fairness as the following optimization problem:
\begin{align*}
\mbox{Max }&\mbox{$\sum_i$} w_i\log r_i\\
=&\mbox{$\sum_i w_i[\log B_{i,n(i),c(n(i))} +
\log\phi_{i,n(i)}+\log\frac{p_{n(i)}}{1-p_{n(i)}}$}\\
&\mbox{$+\log\prod_{m\in\mathcal{M}^{n(i),c(n(i))},
c(m)=c(n(i))}(1-p_m)],$}\\
\mbox{s.t. }&c(n)\in \mathbb{C}, \mbox{ for all } n;\hspace{15pt}n(i)\in \mathcal{N}_i, \mbox{ for all } i;\\
&0\leq p_n\leq 1, \mbox{ for all } n;\hspace{15pt} \phi_{i,n(i)}\geq 0, \mbox{ for all } i;\\
&\mbox{$\sum_{i:n(i)=n}\phi_{i,n}= 1,$} \mbox{ for all }n.
\end{align*}

Based on this formulation, the problem of achieving weighted
proportional fairness consists of four important components, in
increasing order of time scales: First, whenever the AP accesses the
channel, it needs to schedule one client for service. That is, the
AP has to decide the values of $\phi_{i,n}$. Second, in each time
slot, the AP has to decide whether it should access the channel,
which consists of determining the values of $p_n$. Third, each
client needs to decide which AP it should be associated with, i.e.,
deciding $n(i)$. Finally, each AP $n$ needs to choose a channel,
$c(n)$, to operate in. We denote the four problems as
\emph{Scheduling Problem}, \emph{Channel Access Problem},
\emph{Client Association Problem}, and \emph{Channel Selection
Problem}, respectively. Weighted proportional fairness is achieved
by jointly solving the four problems. The problem of achieving
weighted proportional fairness for the distributed scheme can be
formulated similarly and involves three components: the Channel
Access Problem, which chooses $p_i$ for clients, the Client
Association Problem, and the Channel Selection Problem.

Since the overhead for a client to change the AP it is associated
with and for an AP to change the channel it operates in are high,
solutions to the Client Association Problem and the Channel
Selection Problem are updated at a much slower time scale compared
to solutions to the Scheduling Problem and the Channel Access
Problem. Based on this timescale separation, we first study the
solutions to the Scheduling Problem and the Channel Access Problem,
given fixed solutions to the Client Association Problem and the
Channel Selection Problem. We then study the solutions to the Client
Association Problem and the Channel Selection Problem, under the
knowledge of how solutions to the Scheduling Problem and the Channel
Access Problem react. Thus, solutions to the Client Association
Problem and the Channel Selection Problem are indeed joint solutions
to all the four problems, and their optimal solutions achieve
Pareto-optimal weighted proportional fairness. In addition to
solving the four problems, we will show that the solutions naturally
turn into distributed algorithms where each client/ AP makes
decisions based on local knowledge.

\section{The Scheduling Problem and the Channel Access Problem}
\label{section:scheduling contention}

In this section, we assume that solutions to the Client Association
Problem and the Channel Selection Problem, i.e., $n(i)$ and $c(n)$,
are fixed.

Since $n(i)$ and $c(n)$ are fixed, values of $B_{i,n(i),c(n(i))}$
are constant. The optimization problem can be rewritten as
\begin{align*}
\mbox{Max }&\mbox{$\sum_{i\in\mathbb{I}} w_i[\log\phi_{i,n(i)}+\log\frac{p_{n(i)}}{1-p_{n(i)}}$}\\
&\mbox{$+\log\prod_{m\in\mathcal{M}^{n(i),c(n(i))},
c(m)=c(n(i))}(1-p_m)]$}\\
=&\mbox{$\sum_{i\in\mathbb{I}}$} w_i\log\phi_{i,n(i)}+\mbox{$\sum_{n\in\mathbb{N}}$} [w^n\log{p_n}\\
&+(\mbox{$\sum_{m\in\mathcal{M}^{n,c(n)},c(m)=c(n)}$}w^m-w^n)\log(1-p_n)],\\
\mbox{s.t. }&0\leq p_n\leq 1, \mbox{ for all } n;\hspace{15pt} \phi_{i,n(i)}\geq 0, \mbox{ for all } i;\\
&\mbox{$\sum_{i:n(i)=n}$}\phi_{i,n}= 1, \mbox{ for all }n,
\end{align*}
where $w^n=\sum_{i:n(i)=n}w_i$, as defined in Section
\ref{section:system model}. We also define
$z^n:=\sum_{m\in\mathcal{M}^{n,c(n)},c(m)=c(n)}w^m$ to be the total
weights of clients that are associated with APs that interfere with
$n$, including itself.

This formulation naturally decomposes the optimization problem into
two independent parts: maximizing $\sum_i w_i\log\phi_{i,n(i)}$ over
$\phi_{i,n}$, which is the Scheduling Problem, and maximizing
$\sum_n [w^n\log{p_n}+(z^n-w^n)\log(1-p_n)]$ over $p_n$, which is
the Channel Access Problem. Thus, we can solve these two problems
independently.

We first solve the Scheduling Problem.
\begin{theorem}\label{theorem:scheudling contention:scheduling}
Given $n(i)$,
$c(n)$, and $p_n$, for all $i$ and $n$, $\sum_iw_i\log r_i$ is
maximized by setting $\phi_{i,n(i)}\equiv w_i/w^{n(i)}$.
\end{theorem}
\begin{IEEEproof}
We have $\frac{\partial}{\partial \phi_{i,n(i)}}
(\sum_{j\in\mathbb{I}}w_j\log r_j)=\frac{w_i}{\phi_{i,n(i)}}$. Since
$\sum_iw_i\log r_i$ is concave in $[\phi_{i,n}]$ with the condition
$\sum_{i:n(i)=n}\phi_{i,n}= 1$, we have that
$\frac{w_i}{\phi_{i,n}}=\frac{w_j}{\phi_{j,n}}$, for all $i,j$ such
that $n(i)=n(j)=n$, at the optimal point. By setting
$\phi_{i,n(i)}\equiv w_i/w^{n(i)}$, the aforementioned criterion is
satisfied. The conditions $\phi_{i,n(i)}\geq 0, \mbox{ for all } i,$
and $\sum_{i:n(i)=n}\phi_{i,n}= 1, \mbox{ for all }n,$ are also
satisfied. Thus, the Scheduling Problem is solved by setting
$\phi_{i,n(i)}\equiv w_i/w^{n(i)}$.
\end{IEEEproof}

We solve the Channel Access Problem next. The following theorem is
the direct result of Theorem 1 in \cite{gupsto06}.
\begin{theorem}
\label{theorem:scheduling contention:contention} Given $n(i)$,
$c(n)$, and $\phi_{i,n}$, for all $i$ and $n$, $\sum_iw_i\log r_i$
is maximized by setting $p_n\equiv w^n/z^n$.
\end{theorem}

In summary, when the solutions to the Client Association Problem and
the Channel Selection Problem, i.e., $n(i)$ and $c(n)$, are fixed,
the AP $n$ should access the channel with probability $p_n= w^n/z^n$
in each time slot and should schedule the transmission for its
client $i$ with probability $\phi_{i,n}=w_i/w^n$ whenever it
accesses the channel. In addition to achieving the optimal solution
to both the Scheduling Problem and the Channel Access Problem, this
solution only requires $n$ to know the local information of $w^m$
and $c(m)$ for all AP $m$ that may interfere with itself. Thus, this
solution can be easily implemented distributedly.

For the distributed scheme, a theorem similar to Theorem
\ref{theorem:scheduling contention:contention} shows that the
Channel Access Problem is optimally solved by choosing $p_i =
w_i/z^{n(i)}$.

\section{The Client Association Problem and the Channel Selection
Problem}    \label{section:AP channel}

We now propose a distributed algorithm that solves the Client
Association Problem and the Channel Selection Problem based on the
knowledge of optimal solutions to the Scheduling Problem and the
Channel Access Problem. These two problems are non-convex and a
local optimal solution to the two problems may not be globally
optimum, which we will also illustrate by simulations in
Section~\ref{section:simulation}. Thus, common techniques for
solving convex problems are not suitable for these problems.
Instead, the proposed algorithm uses a simulated annealing technique
that is based on the Gibbs Sampler \cite{SG84}, which is proven to
converge to the global optimum point almost surely. We first give an
overview of the technique. We then describe a centralized algorithm
that achieves weighted proportional fairness using the Gibbs
Sampler. Finally, we discuss how to turn this centralized algorithm
into a distributed protocol.

We call a joint solution to both the Client Association Problem and
the Channel Selection Problem as a \emph{configuration} of the
system. A configuration is thus fully specified by the AP each
client is associated with, and the channel each AP operates in.
Define $\psi_t$ as the configuration of the system at time $t$. We
define the \emph{energy} of the system under configuration $\psi_t$,
which we denote by $U(\psi_t)$, as the value of $\sum_i w_i\log r_i$
when APs and clients choose their channels to operate in and APs to
be associated with according to $\psi_t$, and apply the optimal
solution to the Scheduling Problem and the Channel Access Problem
under $\psi_t$. We then have
\begin{equation}
\begin{array}{rl}
U(\psi_t)=&\sum_{i\in\mathbb{I}} w_i[\log
B_{i,n(i),c(n(i))}+\log\frac{w_i}{w^{n(i)}}]\\
&+\sum_{n\in\mathbb{N}}
[w^n\log\frac{w^n}{z^n}+(z^n-w^n)\log\frac{z^n-w^n}{z^n}]
\end{array}
\end{equation}
Finding the joint solution that achieves Pareto-optimal proportional
fairness is equivalent to finding the configuration $\psi$ that
maximizes $U(\psi)$.

We apply the Gibbs sampler to solve the Client Association Problem
and the Channel Selection Problem jointly. At each time $t$, either
a client or an AP is selected according to some arbitrary sequence.
The selected client, or AP, then changes the AP it is associated
with, or the channel it operates in, randomly, while all other
clients and APs make no changes. The solutions to the Scheduling
Problem and the Channel Access Problem are then updated according to
the new configuration.

We now discuss how the selected client, or AP, changes the AP it is
associated with, or the channel it operates in. Let $\psi_t(n(i)=n)$
be the configuration where client $i$ is associated with AP $n$, and
the remaining of the system is the same as in configuration
$\psi_t$. We can define $\psi_t(c(n)=c)$ for AP $n$ similarly. If
client $i$ is selected at time $t$, it changes the AP it is
associated with to $n$ with probability
$e^{U(\psi_t(n(i)=n))/T(t)}/\sum_m e^{U(\psi_t(n(i)=m))/T(t)}$,
where $T(t)$ is a positive decreasing function. On the other hand,
if AP $n$ is selected at time $t$, it changes the channel it
operates in to $c$ with probability
$e^{U(\psi_t(c(n)=c))/T(t)}/\sum_d e^{U(\psi_t(c(n)=d))/T(t)}$.

\cite{SG84} proves that this simple randomized approach maximizes
$U(\psi)$.

\begin{theorem} \label{theorem:AP channel:gibbs}
If $T(t)$ satisfies the following conditions:
\begin{enumerate}
\item $T(t)\rightarrow 0$, as $t\rightarrow\infty$;
\item $T(t)\log t\rightarrow \infty$, as $t\rightarrow \infty$;
\end{enumerate}
then $\lim_{t\rightarrow \infty}U(\psi_t)=\max_{\psi}U(\psi)$ with
probability 1, for any initial configuration $\psi_1$.
\end{theorem}

It remains to compute the values of $U(\psi_t(n(i)=n))$ for client
$i$ and $U(\psi_t(c(n)=c))$ for AP $n$. We first discuss how to
compute $U(\psi_t(n(i)=n))$. Let $w^n_{-i}:=\sum_{j:n(j)=n, j\neq
i}w_j$ be the total weights of clients, excluding $i$, associated
with AP $n$.  Let
$z^n_{-i}:=\sum_{m\in\mathcal{M}^{n,c(n)},c(m)=c(n)}w^m_{-i}$.
Define
\begin{align*}
U^0_i(\psi_t)=&\sum_{j\in\mathbb{I},j\neq i} w_j[\log
B_{j,n(j),c(n(j))}+\log\frac{w_j}{w^{n(j)}_{-i}}]\\
&+\sum_{n\in\mathbb{N}}
[w^n_{-i}\log\frac{w^n_{-i}}{z^n_{-i}}+(z^n_{-i}-w^n_{-i})\log\frac{z^n_{-i}-w^n_{-i}}{z^n_{-i}}]
\end{align*}
which can be thought of as the energy of the system as if the weight
of client $i$ were zero. We then define $\Delta
U^n_{i}(\psi_t):=U(\psi_t(n(i)=n))-U^0_{i}$. Since in the
configuration $\psi_t(n(i)=n)$, $w^m=w^m_{-i}$ for all $m\neq n$;
$w^n=w^n_{-i}+w_i$; $z^m=z^m_{-i}+w_i$ if $m\in\mathcal{M}^{n,c(n)}$
, $c(m)=c(n)$, and $m\neq n$; and $z^m=z^m_{-i}$, otherwise, we have
\begin{align*}
&\Delta U^n_{i}(\psi_t)\\
&=\mbox{$w_i[\log
B_{i,n,c(n)}+\log\frac{w_i}{w^n}]+\sum_{j:n(j)=n}w_j\log\frac{w^n_{-i}}{w^n_{-i}+w_i}$}\\
&\mbox{$+w_i\log\frac{w^n}{z^n}+w^n_{-i}\log\frac{z^n_{-i}(w^n_{-i}+w_i)}{w^n_{-i}(z^n_{-i}+w_i)}+(z^n_{-i}-w^n_{-i})\log\frac{z^n_{-i}}{z^n_{-i}+w_i}$}\\
&\mbox{$+\sum_{m\in\mathcal{M}^{n,c(n)},m\neq n,c(m)=c(n)}[w^m_{-i}\log\frac{z^m_{-i}}{z^m_{-i}+w_i}$}\\
&\mbox{$(z^m_{-i}-w^m_{-i})\log\frac{(z^m_{-i}-w^m_{-i}+w_i)z^m_{-i}}{(z^m_{-i}+w_i)(z^m_{-i}-w^m_{-i})}+w_i\log\frac{z^m-w^m}{z^m}]$}\\
&=\mbox{$w_i[\log\frac{B_{i,n,c(n)}w_i}{z^n}+\sum_{\begin{array}{l}\hspace{-5pt}_{m\in\mathcal{M}^{n,c(n)},}\\
\hspace{-5pt}^{m\neq n,c(m)=c(n)}\end{array}}\hspace{-10pt}\log\frac{z^m-w^m}{z^m}]$}\\
&\mbox{$+\sum_{\begin{array}{l}\hspace{-5pt}_{m\in\mathcal{M}^{n,c(n)},}\\
\hspace{-5pt}^{m\neq n,c(m)=c(n)}\end{array}}\hspace{-10pt}\log[(1+\frac{w_i}{z^m_{-i}-w^m_{-i}})^{z^m_{-i}-w^m_{-i}}/(1+\frac{w_i}{z^m_{-i}})^{z^m_{-i}}]$}\\
&\mbox{$+\log(1-\frac{w_i}{z^n_{-i}+w_i})^{z^n_{-i}}$}\\
&\approx\mbox{$w_i[\log(\frac{B_{i,n,c(n)}w_i}{z^n}\prod_{\begin{array}{l}\hspace{-5pt}_{m\in\mathcal{M}^{n,c(n)},}\\
\hspace{-5pt}^{m\neq
n,c(m)=c(n)}\end{array}}\hspace{-10pt}\frac{z^m-w^m}{z^m})]+\alpha$},
\end{align*}
where $\alpha$ is a constant. Since $(1+\frac{w_i}{A})^A\approx
e^{w_i}$ and $(1-\frac{w_i}{A})^A\approx e^{-w_i}$ for all $A>>w_i$,
the last approximation holds when $z^m_{-i}>>w_i$, which is true in
a dense network where the weights of all clients are within the same
order.

Suppose a client $i$ is selected to change its state at time $t$, at
which time the configuration of the system is $\psi_t$. The
probability that $i$ chooses AP $n$ to be associated with is
$e^{[U^0_i(\psi_t)+\Delta U^n_i(\psi_t)]/T(t)}/\sum_m
e^{[U^0_i(\psi_t)+\Delta U^m_i(\psi_t)]/T(t)}\approx (\frac{B_{i,n,c(n)}w_i}{z^n}\prod_{\begin{array}{l}\hspace{-5pt}_{m\in\mathcal{M}^{n,c(n)},}\\
\hspace{-5pt}^{m\neq
n,c(m)=c(n)}\end{array}}\hspace{-10pt}\frac{z^m-w^m}{z^m})^{w_i/T(t)}/\gamma$,
where $\gamma$ is the normalizer.

To compute the probability of choosing AP $n$ to be associated with,
client $i$ only needs the values of $B_{i,n,c(n)}$ for all
$n\in\mathcal{N}_i$, $w^m$ and $z^m$ for all $m\in\mathcal{M}_i$.
Thus, this probability can be computed by client $i$ using its local
information. We also note that this probability has the following
properties: First, it increases with $B_{i,n,c(n)}$, meaning that
client $i$ tends to choose the AP that has higher data rate; Second,
it decreases with $z^n$, which is the total weights of clients that
interfere with $n$; Finally, it increases with
$\prod_{m\in\mathcal{M}^{n,c(n)},m\neq
n,c(m)=c(n)}\frac{z^m-w^m}{z^m}$, which is the probability that none
of the APs that interfere with $n$ access the channel in a time
slot. Thus, this probability jointly considers the three important
factors for the Client Association Problem: data rate, interference,
and channel congestion.

Next we discuss the computation of the probability that an AP $n$
should choose channel $c$ to operate in, if it is selected. Let
$z^m_{-n}:=\sum_{o\in\mathcal{M}^{m,c(m)}, c(o)=c(m), o\neq
n}w^o-w^m$. Let $U_n^{0}(\psi_t)$ be the energy of the system under
configuration $\psi_t$, if the weights of all its clients were zero.
That is,
\begin{align*}
&\mbox{$U^0_n(\psi_t)=$}\mbox{$\sum_{j\in\mathbb{I},n(j)\neq n}
w_j[\log
B_{j,n(j),c(n(j))}+\log\frac{w_j}{w^{n(j)}}]$}\\
&\mbox{$+\sum_{m\in\mathbb{N},m\neq n}
[w^m\log\frac{w^m}{z^m_{-n}}+(z^m_{-n}-w^m)\log\frac{z^m_{-n}-w^m}{z^m_{-n}}]$}
\end{align*}
We then define $\Delta U^c_{n}(\psi_t):=U(\psi_t(c(n)=c))-U^0_{n}$.
Since in the configuration $\psi_t(c(n)=c)$, $z^m=z^m_{-n}+w^n$ if
$m\in\mathcal{M}^{n,c(n)}$, $c(m)=c$, and $m\neq n$; and
$z^m=z^m_{-n}$, otherwise, we have
\begin{align*}
&\Delta U^c_{n}(\psi_t)\\
&=\mbox{$\sum_{i:n(i)=n}w_i[\log
B_{i,n,c}+\log\frac{w_i}{w^n}]$}\\
&+\mbox{$w^n\log\frac{w^n}{z^n}+(z^n-w^n)\log\frac{z^n-w^n}{z^n}$}\\
&\mbox{$+\sum_{m\in\mathcal{M}^{n,c(n)},m\neq n,c(m)=c(n)}[w^m\log\frac{z^m_{-n}}{z^m_{-n}+w^n}$}\\
&\mbox{$(z^m_{-n}-w^m)\log\frac{(z^m_{-n}-w^m+w^n)z^m_{-n}}{(z^m_{-n}+w^n)(z^m_{-n}-w^m)}+w^n\log\frac{z^m_{-n}-w^m+w^n}{z^m_{-n}+w^n}]$}.
\end{align*}

When an AP $n$ is selected by the Gibbs sampler at time $t$, it
changes the channel it operates randomly, with the probability of
changing to channel $c$ proportional to
$e^{U(\psi_t(c(n)=c))/T(t)}=e^{[U^0_n(\psi_t)+\Delta
U^c_{n}(\psi_t)]/T(t)}\propto e^{\Delta U^c_{n}(\psi_t)/T(t)}$. We
note that, to compute $\Delta U^c_{n}(\psi_t)$, AP $n$ only needs
the values of $B_{i,n,c}$, $w^i$, for each client $i$ that is
associated with $n$, and $z^m_{-n}$, $w^m$ for all
$m\in\cup_c\mathcal{M}^{n,c}$. Thus, $\Delta U^c_{n}(\psi_t)$ can
also be computed using only local information.

Based on the above discussion, it is straightforward to design a
distributed protocol (DP) using the Gibbs sampler. DP achieves the
Pareto-optimal proportional fairness as $t\rightarrow\infty$ almost
surely. Further, in DP, all clients and APs in the system only need
to exchange information within their local areas, as they only need
local information to compute the probability of choosing an AP to be
associated with or a channel to operate in. Thus, DP is easily
scalable.

In addition to DP, we can also consider a greedy policy (Greedy)
that is easier to implement. Greedy works similar to DP, except that
when a client $i$, or an AP $n$, is selected by the sampler, it
chooses the AP that maximizes $U(\psi_t(n(i)=n))$ to be associated
with, or the channel that maximizes $U(\psi_t(c(n)=c))$ to operate
in, respectively. It is essentially a steepest descent direction
approach and is guaranteed to converge to a local optimal point. In
addition to simple implementation, Greedy is also consistent with
the selfish behavior of clients. Each client $i$ chooses the AP $n$
that maximizes
$\frac{B_{i,n,c(n)}}{z^n}\prod_{o\in\mathcal{M}^{n,c(n)},c(o)=c(n),o\neq
n}\frac{z^o}{z^o+w^o}$, which is indeed the value of $r_i$ when $i$
is associated with $n$. Thus, in Greedy, every client always chooses
to associate with the AP that offers the highest throughput.

A similar protocol can also be designed for the distributed scenario
where clients, instead of APs, contend for channel access. The
protocol also uses the Gibbs sampler as DP. Define $z^n_{-i}$,
$z^m_{-n}$, $U^0_i(\psi_t)$, $\Delta U^n_i(\psi_t)$,
$U^0_n(\psi_t)$, and $\Delta U^c_n(\psi_t)$ similar as in the
server-centric scheme. We can derive that, for the distributed
protocol,
\begin{align*}
&\mbox{$\Delta U^n_i(\psi_t)$}\\=&\mbox{$w_i\log
(B_{i,n,c(n)}\frac{w_i}{z^n_{-i}+w_i}\prod_{\begin{array}{l}\hspace{-5pt}_{j:n(j)\in\mathcal{M}^{n,c(n)},}\\
\hspace{-5pt}^{c(n(j))=c(n),j\neq
i}\end{array}}\frac{z^{n(j)}_{-i}-w_j+w_i}{z^{n(j)}_{-i}+w_i})$}\\&\mbox{$-z^n_{-i}\log\frac{z^n_{-i}+w_i}{z^n_{-i}}$}
\end{align*}
and
\begin{align*}
&\mbox{$\Delta U^c_n(\psi_t)$}\\
=&\mbox{$\sum_{i:n(i)=n}w_i\log B_{i,n,c}\frac{w_i}{z^n}+\sum_{\begin{array}{l}\hspace{-5pt}_{m\in\mathcal{M}^{n,c(n)},}\\
\hspace{-5pt}^{c(m)=c(n),m\neq
n}\end{array}}\hspace{-10pt}w^m\log\frac{z^m_{-n}}{z^m_{-n}+w^n}$}\\
&\mbox{$+\sum_{i:n(i)=n}(z^n-w_i)\log\frac{z^n-w_i}{z^n}.$}
\end{align*}

\section{Simulation Results}    \label{section:simulation}

We have implemented both DP and Greedy algorithms. We also compare
their performances against other state-of-the-art solutions. We only
present simulation results for the server-centric scheme due to
space limitations.

We first introduce the model of channel characteristics in our
simulation. We use the ITU path loss model \cite{ITU} to compute the
received signal strength between two devices. If two devices operate
in a band with frequency $f_c$ and are $d$ meters apart, the
received signal strength of a device by the other is proportional to
$\frac{1}{f_c^2d^\alpha}$, where $\alpha$ is the path loss
coefficient and is set to be 3.5.

We adopt the simulation settings in \cite{lipalyan08}, which models
802.11b channels, as the base case and compute the characteristics
of other channels accordingly. A 802.11b channel operates in the 2.4
GHz band with bandwidth 22 MHz. The bit rate between a client and an
AP is 11 Mbps if the distance between them is within 50 meters, 5.5
Mpbs within 80 meters, 2 Mpbs within 120 meters, and 1 Mps within
150 meters. The \emph{maximum transmission range} of 802.11b
channels is hence 150 meters. We assume that two APs interfere with
each other if the received signal strength of one AP by the other is
above the carrier sense threshold, which, as the settings in ns-2
simulator, is set to be 23.42 times smaller than the received signal
strength at a distance of the maximum transmission range. Using the
ITU path loss model, two APs interfere with each other if they are
within the \emph{interference range}, which is
$150\times(23.42)^{\frac{1}{3.5}} = 369$ meters for 802.11b
channels.

For channels other than 802.11b channels, we assume that each of
these channels can support four different data rates, corresponding
to the four data rates of 802.11b channels. Values of each supported
data rate is proportional to the bandwidth of the channel. The
transmission ranges of each data rate and the interference range are
computed so that the received signal strengths at the boundary of
each range is the same to that at the boundary of its counterpart in
802.11b channels. For example, consider a channel that operates in
frequency 4 GHz with bandwidth 44 MHz. The bit rate between a client
and an AP is $11\times \frac{44}{22} = 22$ Mbps if the distance
between them is within $50/(\frac{4}{2.4})^{\frac{2}{3.5}}=37.34$
meters, 11 Mbps within 59.75 meters, 4 Mpbs within 89.62 meters, and
2 Mbps within 112.03 meters. The interference range of this channel
is 275.59 meters.

We compare our algorithms, DP and Greedy, against policies that use
state-of-the-art techniques for solving the Client Association
Problem and the Channel Selection Problem. We compare with
\cite{kaubaccha07}, which proposes a distributed algorithm for
achieving minimum total interference among APs, for the Channel
Selection Problem. For the Client Association Problem and the
Scheduling Problem, we compare with two techniques. The first
technique uses a Wifi-like approach where clients are associated
with the closest AP and the AP schedules clients so that the
throughput of each client is the same. The protocol that applies
both \cite{kaubaccha07} and the Wifi-like approach is called
\emph{MinInt-Wifi}. The other technique is one that is proposed in
\cite{lipalyan08}, which, under a fixed solution of the Channel
Selection Problem, is a centralized algorithm that aims to find the
joint optimal solution to the Client Association Problem and the
Scheduling Problem that achieves weighted proportional fairness.
This technique first relaxes the Client Association Problem by
assuming that each client can be associated with more than one APs
and formulates the problem as a convex programming problem. It then
rounds up the solution to the convex programming problem and finds a
solution to the Client Association Problem where each client is
associated with only one AP. For ease of comparison, we use the
solutions to the relaxed convex programming problem, which is indeed
an upper-bound on the performance of \cite{lipalyan08}. The protocol
that applies both \cite{kaubaccha07} and \cite{lipalyan08} is called
\emph{MinInt-PF}. The Channel Access Problem is then solved by the
optimal solutions based on the resulting solutions of MinInt-Wifi
and MinInt-PF, respectively.

In each of the following simulations, we initiate the system by
randomly assigning channels to each radio of APs. Each client is
initially associated with the closest radio, with ties broken
randomly. The system then evolves according to the evaluated
policies. We compare the policies on two metrics: the weighted sum
of the logarithms of throughput for clients,
$\sum_{i\in\mathbb{I}}w_i\log r_i$, and the total weighted
throughput $\sum_{i\in\mathbb{I}}w_i r_i$. All reported data are the
average over 20 runs. We first show the simulation results for a
simple system that consists of 3 APs and 2 different channels. While
this system may be simplistic, it offers insights on the behavior of
each policy. We then show the simulation results for a larger system
where the list of available channels is gathered from a real-world
scenario.

We first consider a system with 3 APs, each with one radio, that are
separated by 75 meters and are located at positions (0,0), (75,0),
and (150,0). There are 16 clients, all with weights 1.0, and the
$i^{th}$ client is located at position $(35+5i,0)$. We consider two
settings for channels: one that with only one 802.11b channel and
the other with one 802.11b channel and a channel that operates at
frequency 16 GHz with bandwidth 50 MHz. This channel can support
higher data rates but has smaller transmission and interference
ranges.

\begin{figure}[t]\subfloat{
\label{fig:pf} 
\includegraphics[width=1.6in]{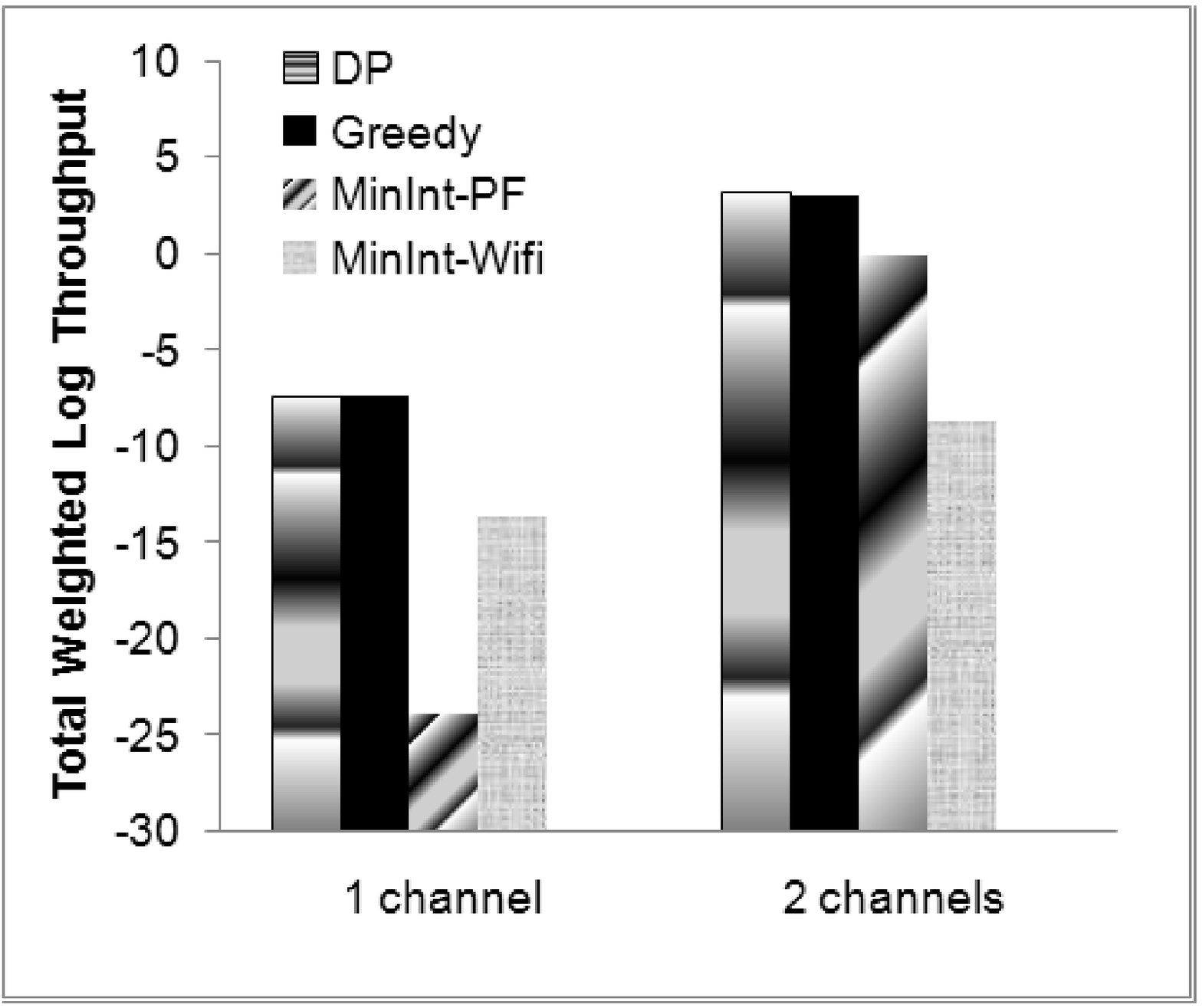}}
\hspace{0.01\linewidth}\subfloat{
\label{fig:throughput} 
\includegraphics[width=1.6in]{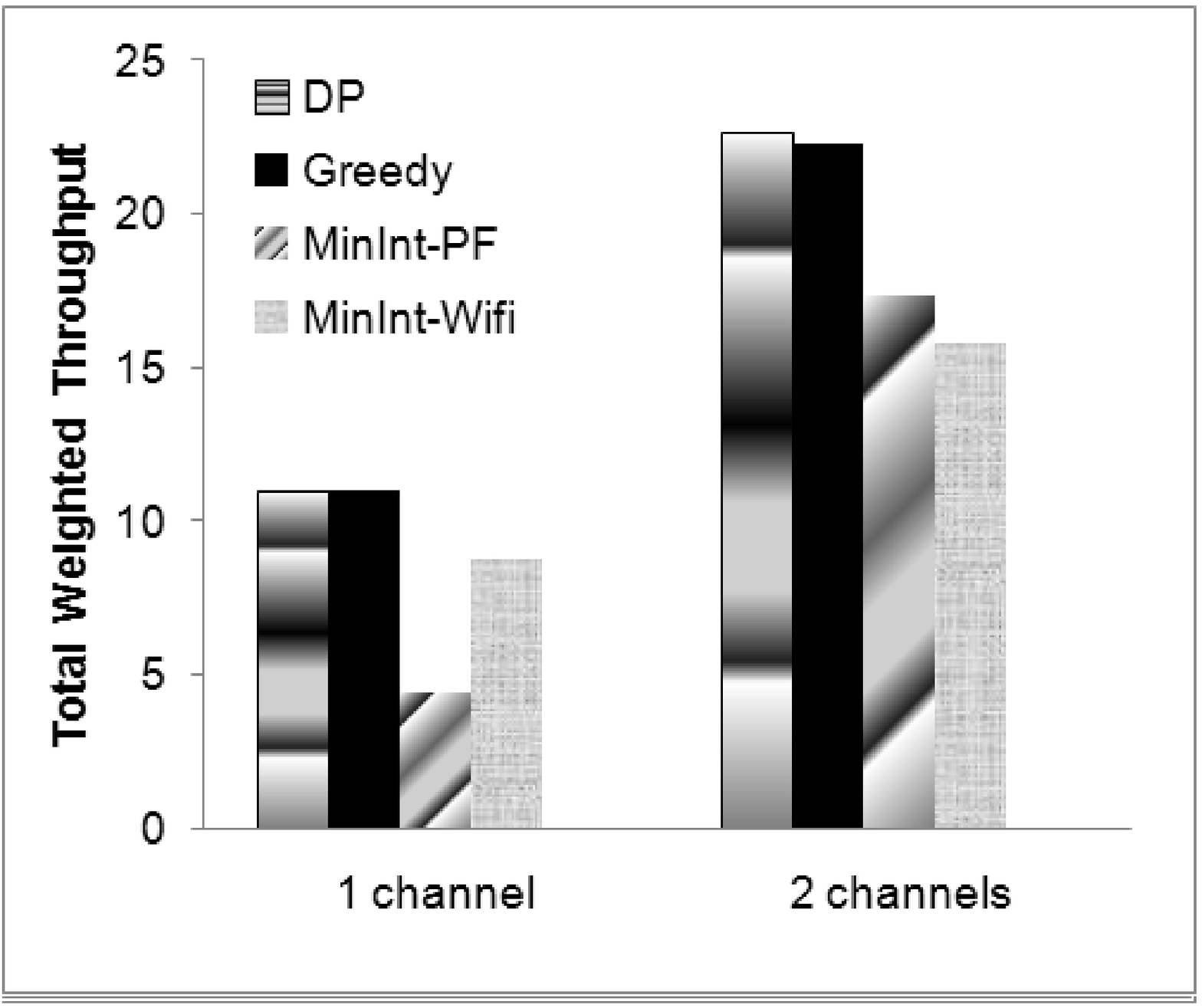}}
\caption{Performance comparison for a simple
system.}\label{fig:performance}
\end{figure}

Simulation results are shown in Fig. \ref{fig:performance}. DP and
Greedy outperform MinInt-Wifi and MinInt-PF in both evaluated
metrics under both 1-channel and 2-channel settings. For the case
where there is only one channel, the solutions of the Channel
Selection Problem does not have any influence on the results.
MinInt-PF does not have good performance because it distribute
clients equally to all three APs, which leads to serious contention
and collisions within the network. MinInt-Wifi also suffers from the
same problem. On the other hand, under DP, all clients are
associated with the AP located at (75,0) and therefore contention is
avoided. This result suggests that a desirable algorithm for the
Client Association Problem also needs to jointly consider the
effects on both the Scheduling Problem and the Channel Access
Problem.

For the case where there are two channels, both MinInt-Wifi and
MinInt-PF select the APs at (0,0) and (150,0) to operate in the
channel at frequency 16 GHz with bandwidth 50 MHz and the AP at
(75,0) to operate in the 802.11b channel. This selection is the only
one that results in no interference within the network. On the other
hand, DP selects the APs at (0,0) and (150,0) to operate in the
802.11b channel and the AP at (75,0) to operate in the other
channel. While this selection results in interference between the
APs at (0,0) and (150,0), DP actually achieves better performance in
both metrics. This is because, in our setting, most clients are
gathered around the AP at (75,0) and thus an optimal solution should
allow the AP at (75,0) to operate in a channel with higher data
rates. This shows that an algorithm that aims to minimize
interference among APs may not be optimal because it fails to
consider the geographical distribution of clients. Further, although
the performance of Greedy is suboptimal, which is because the Client
Association Problem and the Channel Selection Problem are
non-convex, it is actually close to that of DP and is much better
than those of MinInt-Wifi and MinInt-PF.

Next, we consider a larger system. The system consists of 16 APs
that are planned as a 4 by 4 grid. Each AP has 2 radios and adjacent
APs are separated by 300 meters. There are 16 clients uniformly
distributed in each of the two sectors $[0,300]\times[0,300]$ and
$[600,900]\times[600,900]$; There are 9 clients uniformly
distributed in each of the two sectors $[0,300]\times[600,900]$ and
$[600,900]\times[0,300]$. We consider the TV white spaces available
in New York City \cite{debgupkan10}. The list of available channels
is shown in Table \ref{table:simulation:nyc}. We consider two
settings: an unweighted setting where all clients have weights 1.0,
and a weighted setting where clients within the region
$[0,300]\times[0,900]$ have weights 1.5 and clients outside this
region have weights 0.5.

\begin{figure}[t]\subfloat{
\label{fig:pf} 
\includegraphics[width=1.6in]{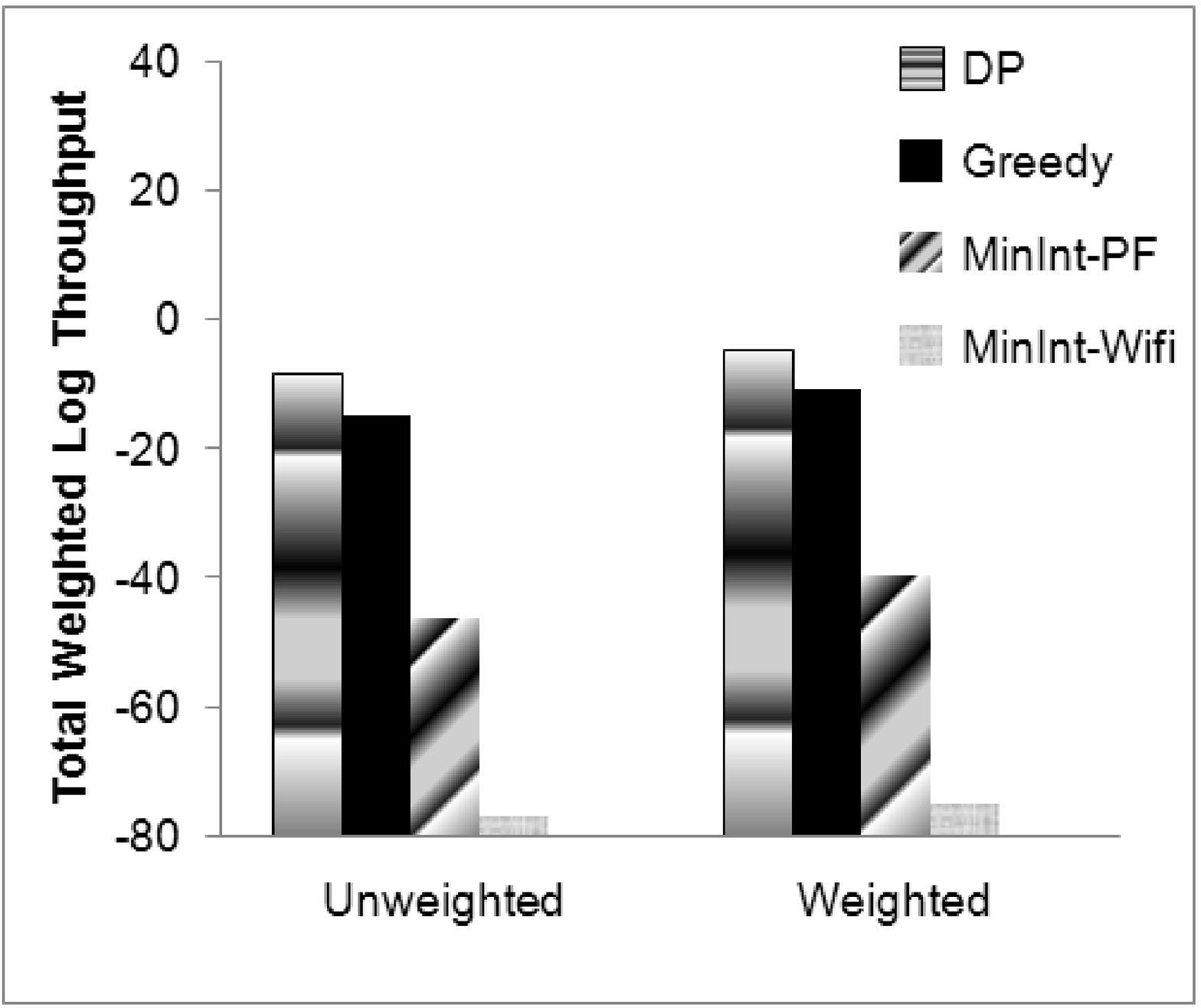}}
\hspace{0.01\linewidth}\subfloat{
\label{fig:throughput} 
\includegraphics[width=1.6in]{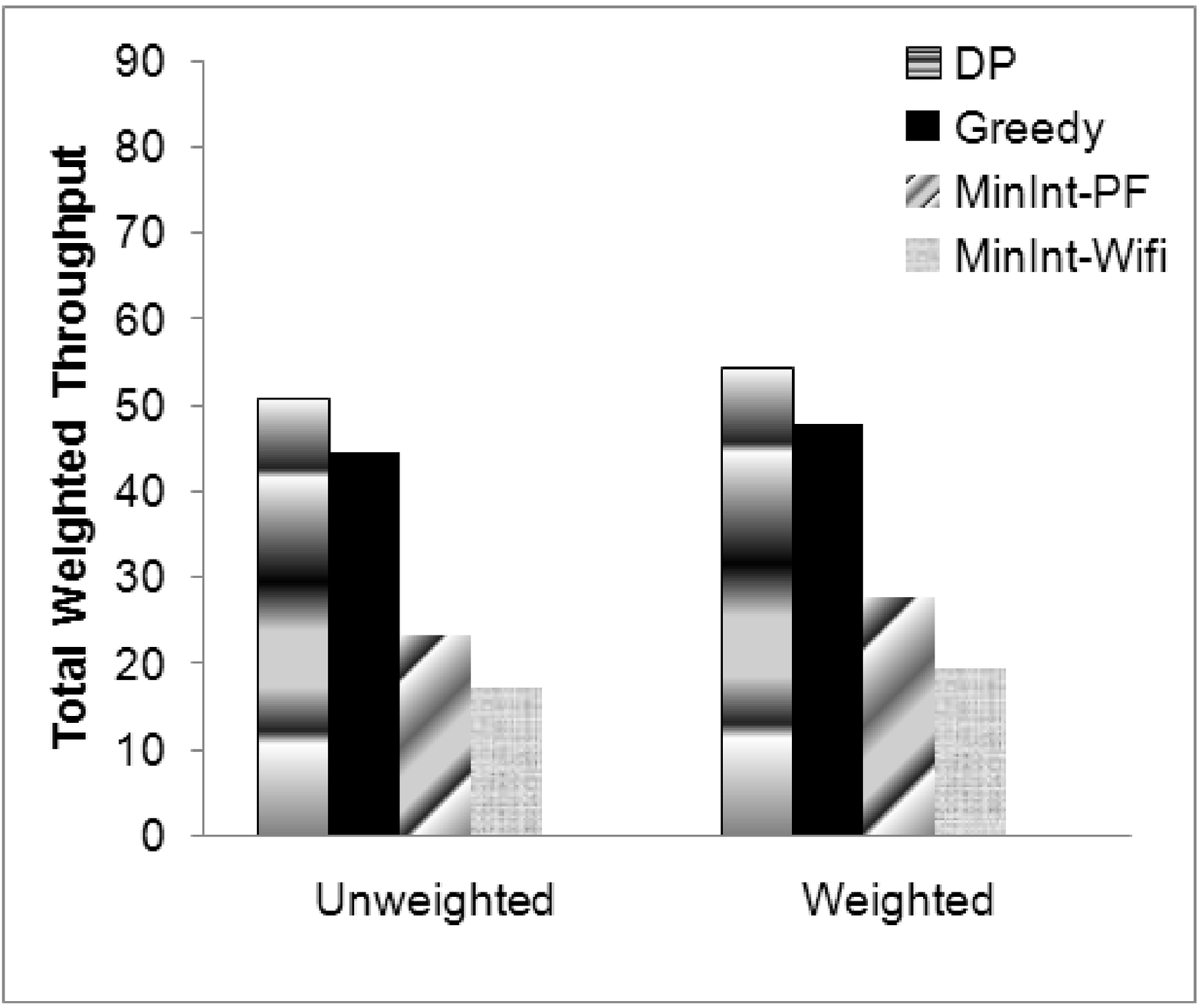}}
\caption{Performance comparison for a larger
system.}\label{fig:performance2}
\end{figure}

\begin{table}[tb]
\begin{center}
\begin{tabular}{|c|c|c||c|c|c|}
\hline
id & frequency & bandwidth & id & frequency & bandwith\\
\hline A &  524 MHz & 12 MHz & E & 659 MHz & 6 MHz\\
\hline B & 593 MHz & 6 MHz & F & 671 MHz & 6 MHz\\
\hline C & 608 MHz & 12 MHz & G & 683 MHz & 6 MHz\\
\hline D & 641 MHz & 6 MHz & & &\\
\hline
\end{tabular}
\end{center}
\caption{List of white spaces in New York City.}
\label{table:simulation:nyc}
\end{table}
Simulation results are shown in Fig. \ref{fig:performance2}. For
both the unweighted and weighted settings, MinInt-Wifi and MinInt-PF
are far from optimum. The total weighted throughputs achieved by the
two policies are less than half of those achieved by DP under both
settings. The performance of Greedy is close to optimum, whose
weighted total throughputs are about 85$\%$ and 84$\%$ of those by
DP for the unweighted and weighted settings, respectively.

\section{Conclusion}    \label{section:conclusion}

We have studied the problem of achieving weighted proportional
fairness in multi-band wireless networks. We have considered a
system that consists of several APs and clients operating in a
number of available channels, accounting for interference among APs
and heterogeneous characteristics of different channels. We have
identified that the problem of achieving weighted proportional
fairness in such a system involves four important components: client
scheduling, channel access, client association, and channel
selection. We have proposed a distributed protocol that jointly
considers the four components and achieves weighted proportional
fairness. We have also derived a greedy policy based on the
distributed protocol that is easier to implement. Simulation results
have shown that the distributed protocol outperforms
state-of-the-art techniques. The total weighted throughputs achieved
by the distributed protocol can be twice as large as
state-of-the-art techniques. Simulation results have also shown
that, while being suboptimal, the performance of the greedy policy
is actually close to optimum quite often.

\def\baselinestretch{0.85}
\bibliographystyle{plain}
\bibliography{reference}
\def\baselinestretch{0.84}
\normalsize
\end{document}